\pdfoutput=1
\documentclass[11pt]{article}
\usepackage{array}
\usepackage{longtable}

\usepackage[final]{acl}

\usepackage{times}
\usepackage{latexsym}
\usepackage{subcaption} 
\usepackage[T1]{fontenc}
\usepackage[utf8]{inputenc}
\usepackage{afterpage}
\usepackage{microtype}
\usepackage{amsmath} 
\usepackage{hyperref}

\usepackage{inconsolata}

\usepackage{graphicx}
\usepackage{float}
\usepackage{cleveref}
\title{Lla-VAP: LSTM Ensemble of Llama and VAP for Turn-Taking Prediction}

\author{Hyunbae Jeon \and Frederic Guintu \and Rayvant Sahni \\
 Department of Computer Science\\
 Emory University\\
 Atlanta, GA 30322 \\
 \texttt{\{harry.jeon,frederic.guintu,rayvant.sahni\}@emory.edu}
}
\begin{document}
\maketitle
\begin{abstract}
Turn-taking prediction is the task of anticipating when the speaker in a conversation will yield their turn to another speaker to begin speaking. This project expands on existing strategies for turn-taking prediction by employing a multi-modal ensemble approach that integrates large language models (LLMs) and voice activity projection (VAP) models. By combining the linguistic capabilities of LLMs with the temporal precision of VAP models, we aim to improve the accuracy and efficiency of identifying TRPs in both scripted and unscripted conversational scenarios. Our methods are evaluated on the In-Conversation Corpus (ICC) and Coached Conversational Preference Elicitation (CCPE) datasets, highlighting the strengths and limitations of current models while proposing a potentially more robust framework for enhanced prediction.
\end{abstract}

\section{Introduction}
Turn-taking is a fundamental mechanism in human communication, ensuring orderly and coherent verbal exchanges. It involves participants alternating roles as speaker and listener, guided by verbal and non-verbal cues that signal when a speaker's turn is nearing its end. These signals create opportunities, known as Transition Relevance Places (TRPs), where the conversational floor can shift to another participant. TRPs are not fixed but dynamically emerge from the interplay of syntactic, semantic, prosodic, and contextual cues within a conversation \cite{umair2024llm}.

The ability to predict TRPs is important in the process of designing effective spoken dialogue systems (SDS) and conversational agents. Human interlocutors anticipate TRPs to avoid interruptions, minimize silence, and maintain conversational flow. However, artificial systems often struggle with this task, particularly in unscripted interactions where the timing and nature of TRPs vary significantly. Current models like TurnGPT and RC-TurnGPT demonstrate the potential of large language models (LLMs) for turn-taking tasks but remain limited to text-based features, lacking integration of acoustic cues critical for real-world conversations \cite{ekstedt-skantze-2020-turngpt}.

Our project addresses these limitations by proposing a multi-modal approach that leverages both linguistic and acoustic features. Specifically, we integrate a Large Language Model (LLM) with a Voice Activity Projection (VAP) model to capitalize on their complementary strengths. LLMs excel in understanding the syntactic and semantic context of conversations, while VAP models provide insights into non-verbal cues, such as pauses and tonal shifts, that signal TRPs. This combination aims to enhance the accuracy and robustness of turn-taking prediction across diverse conversational settings \cite{wang2024turntakingbackchannelpredictionacoustic}.

Building on prior research, we utilize the In-Conversation Corpus (ICC) as a primary dataset to explore unscripted turn-taking dynamics. The ICC, with its participant-labeled within-turn TRPs, provides a nuanced understanding of conversational behavior and highlights the challenges LLMs face in predicting these nuanced interaction points. Unlike scripted datasets, the ICC reveals the variability and subtlety of natural conversations, making it an ideal testbed for our proposed methodology. Complementing this, we employ the Coached Conversational Preference Elicitation (CCPE) dataset to analyze turn-taking in controlled task-oriented dialogues, facilitating a comprehensive evaluation of our models \cite{radlinski-etal-2019-coached}.

By bridging the gap between single-mode and multi-modal approaches, our research seeks to advance the state of turn-taking prediction. Beyond its technical implications, accurate TRP prediction holds broader significance for improving human-computer interaction, enabling conversational agents to engage in dialogues that feel more natural, fluid, and human-like.

% Turn-taking is a fundamental concept in conversation management and discourse, where each participant in the dialogue takes turns speaking, typically one at a time. Alternating turns are determined by both verbal and nonverbal cues given by the current speaker indicating the end of their turn, yielding to another speaker \cite{article}.

% Our research project aims to bridge the gap in single-mode turn-taking prediction by leveraging a  multi-modal approach that combines both text-based and audio-based cues. By integrating a Large  Language Model (LLM) with a Voice Activity Projection (VAP) model, we capitalize on the linguistic  strengths of the LLM in understanding conversational flow and the VAP model’s ability to interpret  non-verbal audio cues, such as pauses and tonal shifts. This dual approach seeks to improve the  accuracy and efficiency of existing turn-taking prediction methods.

\section{Related Work}

With advancements in natural language processing came large language models (LLMs) which are  computational systems that have demonstrated the ability to understand human language and its  contextual nuances. These models can capture linguistic patterns that indicate conversational flow,  as demonstrated by systems like TurnGPT, which is based on a transformer architecture fine-tuned  for turn-taking prediction. TurnGPT effectively identifies turn-shift points by leveraging syntactic  and pragmatic completeness cues, making it a valuable model for turn-prediction tasks \cite{ekstedt-skantze-2020-turngpt}. However,  the TurnGPT model’s understanding remains limited to text-based inputs, thus lacking integration of  nonverbal cues critical for natural conversation dynamics.

The Voice Activity Projection (VAP) model is a continuous turn-taking prediction system designed to  anticipate future voice activity in conversations, contrasting traditional models that rely on simple  silence thresholds. The VAP model uses multi-layer transformers to process raw audio signals and  predict the likelihood of voice activity for each speaker over a specified time frame. This approach  allows the model to detect subtle cues in conversational dynamics, such as pauses and shifts, essential  for turn-taking decisions \cite{inoue2024realtimecontinuousturntakingprediction}.

Recently, Wang et al. \cite{wang2024turntakingbackchannelpredictionacoustic} proposed an method to turn-taking prediction by combining a neural  acoustic model with a large language model. Their experiments demonstrated that this multi-modal fusion of lexical and acoustic information "consistently outperforms baseline models with single modality," highlighting that an effective integration of both linguistic and audio cues enhances predictive accuracy. These findings indicate the potential of multi-modal approaches to significantly improve outcomes in turn-taking prediction when properly implemented \cite{wang2024turntakingbackchannelpredictionacoustic}.

\begin{table*}[ht]
    \centering
    \caption{Properties of CCPE and ICC dataset}
    \resizebox{\textwidth}{!}{%
    \renewcommand{\arraystretch}{1} % Adjust row height
    \begin{tabular}{|p{3cm}|p{6cm}|p{6cm}|}
        \hline
        \textbf{Aspect} & \textbf{CCPE} & \textbf{ICC} \\
        \hline
        Data Composition
        & 502 English dialogues with 12,000 annotated utterances
        & Approximately 93 conversations, each lasting about 25 minutes \\
        \hline 
        Format
        & JSON format
        & High-quality audio recordings \\
        \hline 
        Language
        & English
        & American English \\
        \hline 
        Source
        & Google LLC
        & Human Interaction Lab (HiLab) at Tufts University \\
        \hline 
        Purpose
        & Elicit conversational preferences about movies in a controlled setting, minimizing bias in user language
        & Study informal, unscripted dialogues to analyze turn-taking and identify multiple Turn Construction Units (TCUs) \\
        \hline
    \end{tabular}%
    }
\end{table*}

\section{Dataset}

For this project, we use two primary datasets: the \textit{Coached Conversational Preference Elicitation (CCPE) dataset} \cite{radlinski-etal-2019-coached} and the\textit{ In-Conversation Corpus (ICC)} \cite{umair2024llm}.

\subsection{CCPE}
The Coached Conversational Preference Elicitation is designed specifically for eliciting conversational preferences in a controlled setting. It consists of 502 English dialogues with 12,000 annotated utterances between a user and an assistant discussing movie preferences in natural language. The corpus was constructed from dialogues between two paid crowd-workers using a Wizard-of-Oz methodology. One worker plays the role of an "assistant", while the other plays the role of a "user". The "assistant" is tasked with eliciting the "user" preferences about movies following a Coached Conversational Preference Elicitation (CCPE) methodology. In particular, the assistant is required to ask questions designed so as to minimize the bias in the terminology the "user" employs to convey his or her preferences, and obtain these in as natural language as possible. Each dialog is annotated with entity mentions, preferences expressed about entities, descriptions of entities provided, and other statements of entities. 
The dialogue corpus is provided in JSON format in a file called data.json. \\

\subsection{ICC}
The In-Conversation Corpus consists of high-quality recordings of informal dialogues in American English, collected at the Human Interaction Lab (HiLab) at Tufts University. It features pairs of undergraduate students engaged in unscripted conversations, collected in a controlled environment with sound isolation to ensure clarity. The dataset was filtered to create a subset of turns suspected to contain multiple Turn Construction Units (TCUs), providing a rich source for identifying both turn-final and within-turn TRPs. In total, the ICC consists of approximately 93 conversations, each lasting approximately 25 minutes.

These datasets will be employed to train and evaluate our models, providing diverse conversational scenarios to assess the effectiveness of our turn-taking prediction methods.

\section{Approach}
Our study implements a multi-modal approach to turn-taking prediction that leverages both audio and textual signals. The methodology comprises three main components: an audio-based model, a text-based model, and ensemble methods that combine both modalities.

\subsection{Data Preprocessing}
The effectiveness of our approach relies on careful preprocessing of both datasets:

\paragraph{Audio Processing}
\begin{itemize}
    \item \textbf{CCPE Processing:}
    \begin{itemize}
        \item Generated audio signals from text transcripts using coqui/XTTS-v2
        \item Inserted 2-second artificial silences between turns to simulate TRPs \cite{10.3389/fpsyg.2015.00731}
        \item Applied speaker differentiation through TTS parameters
    \end{itemize}
    
    \item \textbf{ICC Processing:}
    \begin{itemize}
        \item Utilized faster\_whisper small Automatic Speech Recognition (ASR) model for transcription \cite{faster-whisper}
        \item Aligned human-labeled TRPs for stimulus-response pairs
        \item Generated ground truth based on $\geq 30\%$ participant agreement
        \item Adjusted real-time evaluations to account for 1.8-second average prediction window
    \end{itemize}
\end{itemize}

\subsection{Base Models}
Our approach employs two baseline models that capture different aspects of turn-taking:

\begin{itemize}
    \item \textbf{Audio-Based Model}: Utilizes Voice Activity Projection (VAP) to process raw audio waveforms, making incremental predictions about future voice activity. The model processes audio in 1-second chunks \cite{inoue2024realtimecontinuousturntakingprediction} and combines voice activity history with current input.
    
    \item \textbf{Text-Based Model}: Leverages the Llama 3.2 3B Instruct model \cite{grattafiori2024llama3herdmodels} for linguistic understanding, processing transcribed audio segments to predict turn completion points. This approach capitalizes on semantic and syntactic cues in speech.
\end{itemize}

\subsection{Ensemble Strategy}
To capitalize on the complementary strengths of audio and textual features, we developed three ensemble approaches:

\begin{itemize}
    \item \textbf{Logistic Regression}: Combines raw predictions with engineered temporal features
    \item \textbf{Prompt-Based}: Enhances LLM decisions by incorporating VAP confidence scores
    \item \textbf{LSTM}: Captures complex temporal dependencies using bidirectional processing and attention mechanisms
\end{itemize}

\subsection{Evaluation Framework}
Model performance is evaluated using a comprehensive set of metrics that account for both classification accuracy and temporal precision. We employ a $\pm$75 frame evaluation window (1.5 seconds at 50Hz) around ground truth turn-shift points to account for the temporal nature of turn-taking prediction. This window size was chosen based on the average duration of turn shift signals observed in ICC participant responses, while CCPE was preprocessed with 2-second gaps between turns.

Each metric serves a specific purpose in our evaluation:
\begin{itemize}
    \item \textbf{Accuracy} provides an overall measure of correct predictions but can be misleading with imbalanced data
    \item \textbf{Balanced Accuracy} accounts for class imbalance by giving equal weight to positive and negative class performance \\
    \begin{equation}
    \text{Balanced Acc.} = \frac{1}{2} \left(\frac{\text{TP}}{\text{TP} + \text{FN}} + \frac{\text{TN}}{\text{TN} + \text{FP}}\right)
    \end{equation}
    \item \textbf{Precision} measures the proportion of correctly identified turn-taking points among all predicted turns
    \item \textbf{Recall} captures the model's ability to identify all actual turn-taking points
    \item \textbf{F1 Score} provides a balanced measure between precision and recall
    \item \textbf{RTF} assesses computational efficiency for real-time applications \\
    \begin{equation}
    \text{RTF} = \frac{\text{Processing Time}}{\text{Audio Duration}}
    \end{equation}
\end{itemize}

\section{Experiments}

\subsection{Implementation Details}

\subsubsection{Base Models}
\paragraph{Voice Activity Projection (VAP)} 
We implemented VAP following the original design \cite{inoue2024realtimecontinuousturntakingprediction}, processing 1-second audio chunks at 16kHz sampling rate. Threshold optimization was performed by testing both normal and inverted predictions across multiple threshold values.

\paragraph{Llama 3.2 3B Instruct}
We evaluated three distinct prompting strategies: Prompt 1 is based on \citet{umair2024llm}, Prompt 2 focuses on minimizing jargon for better LLM's comprehension, and Prompt 3 leverages the few-shot learning potential of LLMs as described in \cite{ma2023fairnessguidedfewshotpromptinglarge}.

\begin{verbatim}
Prompt 1:
System: You are a conversation 
analysis expert. Your task is to 
identify Transition Relevance 
Places (TRPs) where a listener 
could appropriately take their 
turn speaking. Answer with only 
'yes' or 'no'.
User: Speech segment: {text}

Prompt 2 (Best Performing):
System: You are having a 
conversation with a user. Your 
task is to identify if the user 
has finished their turn. Answer 
with only 'yes' or 'no'.
User: Speech segment: {text}

Prompt 3 (Few-shot):
System: You are having a 
conversation with a user. Your 
task is to identify if the user 
has finished their turn. Answer 
with only 'yes' or 'no'.
Here are some examples:

User: "I was walking in the park"
Assistant: no
User: "I was walking in the park 
last night"
Assistant: yes
User: "I was walking in the park 
last night, and I saw a squirrel."
Assistant: yes
User: "what kind of movie do you 
generally like"
Assistant: yes
User: "I recently actually went 
to see a movie"
Assistant: no

User: Speech segment: {text}
\end{verbatim}

Model parameters were set to temperature=0.1, top-p=0.9, with a maximum context length of 10 seconds.

\subsubsection{Ensemble Implementations}

\paragraph{Logistic Regression Ensemble}
Features included:
\begin{itemize}
    \item Raw probability outputs from both base models
    \item Rolling statistics (5, 10, 20 frame windows):
        \begin{itemize}
            \item Mean and standard deviation
            \item Maximum and minimum values
        \end{itemize}
    \item Interaction features:
        \begin{align}
        P_{combined} = \{ & P_{VAP} \times P_{LLaMA}, \nonumber \\
                         & \max(P_{VAP}, P_{LLaMA}), \nonumber \\
                         & \min(P_{VAP}, P_{LLaMA})\}
        \end{align}
\end{itemize}

\paragraph{Prompt Ensemble}
Enhanced the base Llama prompt with VAP predictions:

\begin{verbatim}
System: You are having a 
conversation with a user. Your 
task is to identify if the user 
has finished their turn. Answer 
with only 'yes' or 'no'.

User: Speech segment: [text]
VAP confidence: [1 - vap_confidence]
(threshold: 0.9, prediction: 
[yes/no based on vap_confidence 
<= 0.1])
\end{verbatim}

\paragraph{LSTM Ensemble}
\begin{itemize}
    \item \textbf{Architecture:} Input(128×2) → BiLSTM(2-layer, 128d) → Attention(4-head) → Dropout(0.3)
    \item \textbf{Training:} Focal($\gamma$=3.0, $\alpha$=0.75) + AdamW(lr=1e-3, wd=0.01) | Batch=32, Seq=100
\end{itemize}

\subsection{Experimental Setup}
For model evaluation, we employed different cross-validation strategies based on dataset characteristics:
\begin{itemize}
    \item CCPE: 5-fold cross-validation
    \item ICC: Leave-one-out cross-validation due to limited stimulus files (4 files)
\end{itemize}

All experiments were conducted on an NVIDIA RTX 3060 Ti 8GB GPU. Real-time factors were measured separately for base models and ensemble methods to accurately reflect computational overhead.

\section{Analysis}

% baseline result sum
\begin{table*}[ht]
    \centering
    \caption{Performance metrics for baseline VAP, Llama, and Whisper models on ICC and CCPE datasets.}
    \label{tab:baseline_results}
    \resizebox{\textwidth}{!}{%
    \begin{tabular}{|c|c|c|c|c|c|c|c|c|c|c|c|}
        \hline
        \textbf{Dataset} & \textbf{Model} & \textbf{Prompt} & \textbf{Threshold} & \textbf{Accuracy} (\%) & \textbf{Balanced Acc} (\%) & \textbf{Precision} & \textbf{Recall} & \textbf{F1 Score} & \textbf{RTF (VAP)} & \textbf{RTF (Llama)} & \textbf{RTF (Whisper)} \\
        \hline
        ICC & VAP & \textit{-} & 0.8 (no flip) & 46.85 & 52.16 & 0.0232 & 0.5770 & 0.0445 & \textbf{0.0203} & - & - \\
        ICC & Llama & Prompt 1 & 0.6 (no flip) & 18.29 & 52.69 & 0.0228 & \textbf{0.8862} & 0.0444 & - & 0.0455 & 0.3220 \\
        ICC & Llama & Prompt 2 & 0.8 (flip) & \textbf{52.86} & \textbf{55.77} & \textbf{0.0259} & 0.5877 & 0.0496 & - & 0.0448 & 0.3227 \\
        ICC & Llama & Prompt 3 & 0.1 (no flip) & 40.67 & 50.00 & 0.4067 & 1.000 & \textbf{0.5693} & - & 0.0610 & 0.3109 \\
        CCPE & VAP & \textit{-} & 0.1 (flip) & \textbf{87.30} & \textbf{86.75} & \textbf{0.8481} & \textbf{0.8313} & \textbf{0.8368} & \textbf{0.0203} & - & - \\
        CCPE & Llama & Prompt 1 & 0.4 (no flip) & 61.89 & 64.23 & 0.5197 & 0.7798 & 0.6167 & - & 0.0414 & 0.3261 \\
        CCPE & Llama & Prompt 2 & 0.75 (no flip) & 79.26 & 77.16 & 0.8300 & 0.6328 & 0.7098 & - & 0.0418 & 0.3023 \\
        CCPE & Llama & Prompt 3 & 0.1 (no flip) & 40.67 & 50.00 & 0.4067 & 1.000 & 0.5693 & - & 0.0550 & 0.3200 \\
        \hline
    \end{tabular}%
    }
\end{table*}

\begin{table*}[htbp]
\centering
\caption{Performance Comparison of Different Ensembles}
\begin{tabular}{lcccccc}
\hline
Method & Accuracy (\%) & Balanced Acc (\%) & Precision & Recall & F1 Score & RTF \\
\hline
IR ICC & 63.53 & 47.15 & 0.002 & 0.307 & 0.004 & 0.0001 \\
Prompt ICC & \textbf{64.00} & 47.50 & 0.002 & 0.310 & 0.004 & N/A \\
LSTM ICC & 12.33 & \textbf{53.80} & 0.002 & \textbf{0.906} & 0.004 & 0.0242 \\
IR CCPE & 78.20 & 78.30 & 0.941 & 0.590 & 0.723 & 0.0001 \\
Prompt CCPE & 82.93 & 79.53 & \textbf{0.982} & 0.600 & 0.742 & N/A \\
LSTM CCPE & \textbf{93.20} & \textbf{96.08} & 0.939 & \textbf{0.991} & \textbf{0.964} & 0.0249 \\
\hline
\end{tabular}
\label{tab:ensemble_results}
\end{table*}

\subsection{Training Dynamics}
\begin{figure*}[t]
    \centering
    \includegraphics[width=\textwidth]{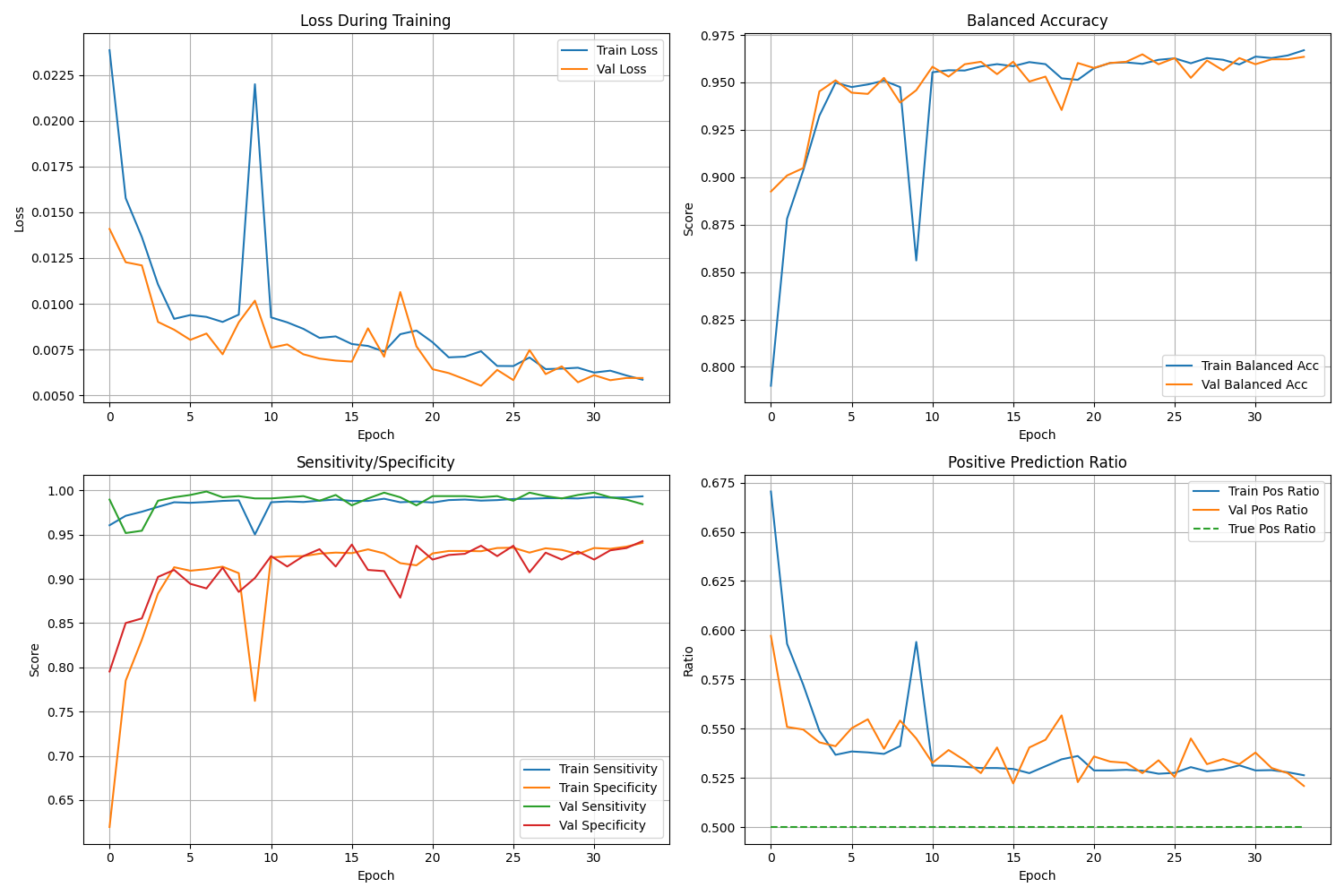}
    \caption{LSTM ensemble training progression on CCPE dataset. (Top-left) Training and validation loss showing convergence. (Top-right) Balanced accuracy achieving and maintaining approximately 95\% after epoch 5. (Bottom-left) Sensitivity/specificity demonstrating balanced learning. (Bottom-right) Positive prediction ratio approaching true distribution (indicated by dashed line).}
    \label{fig:lstm_training}
\end{figure*}

To understand how our LSTM ensemble learns to integrate audio and text features, we analyze its training progression across multiple metrics in Figure \ref{fig:lstm_training}. The four subplots reveal the convergence behavior, accuracy improvements, and prediction characteristics throughout the training process, providing insights into how the model learns to balance different aspects of turn-taking prediction.

\subsection{Model Prediction Analysis}
\begin{figure}[t]
    \centering
    \includegraphics[width=\columnwidth]{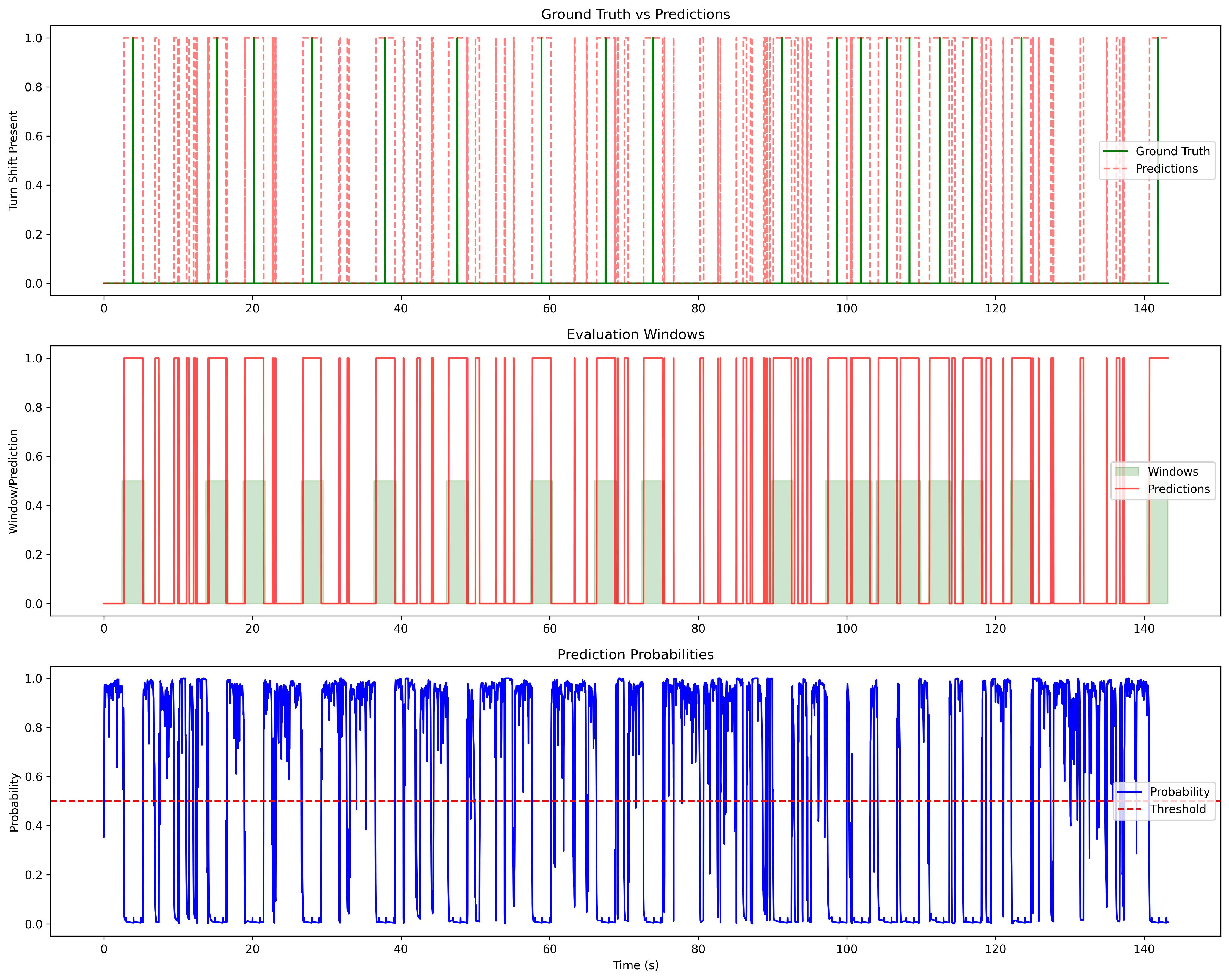}
    \caption{VAP showing less stable predictions}
    \label{fig:VAP_example}
\end{figure}

\begin{figure}[t]
    \centering
    \includegraphics[width=\columnwidth]{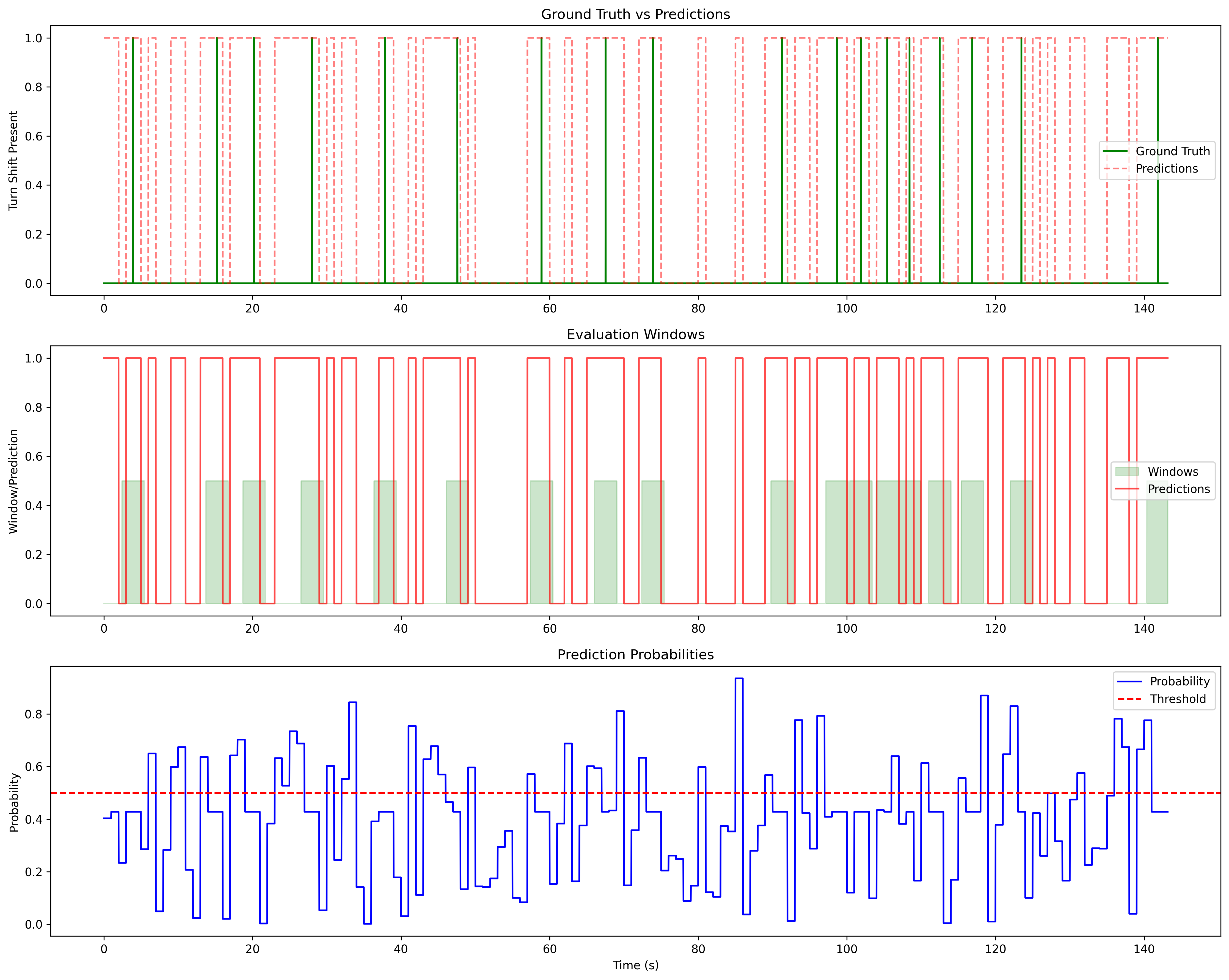}
    \caption{Basic prompt inference showing less stable predictions}
    \label{fig:prompt1_example}
\end{figure}

\begin{figure}[t]
    \centering
    \includegraphics[width=\columnwidth]{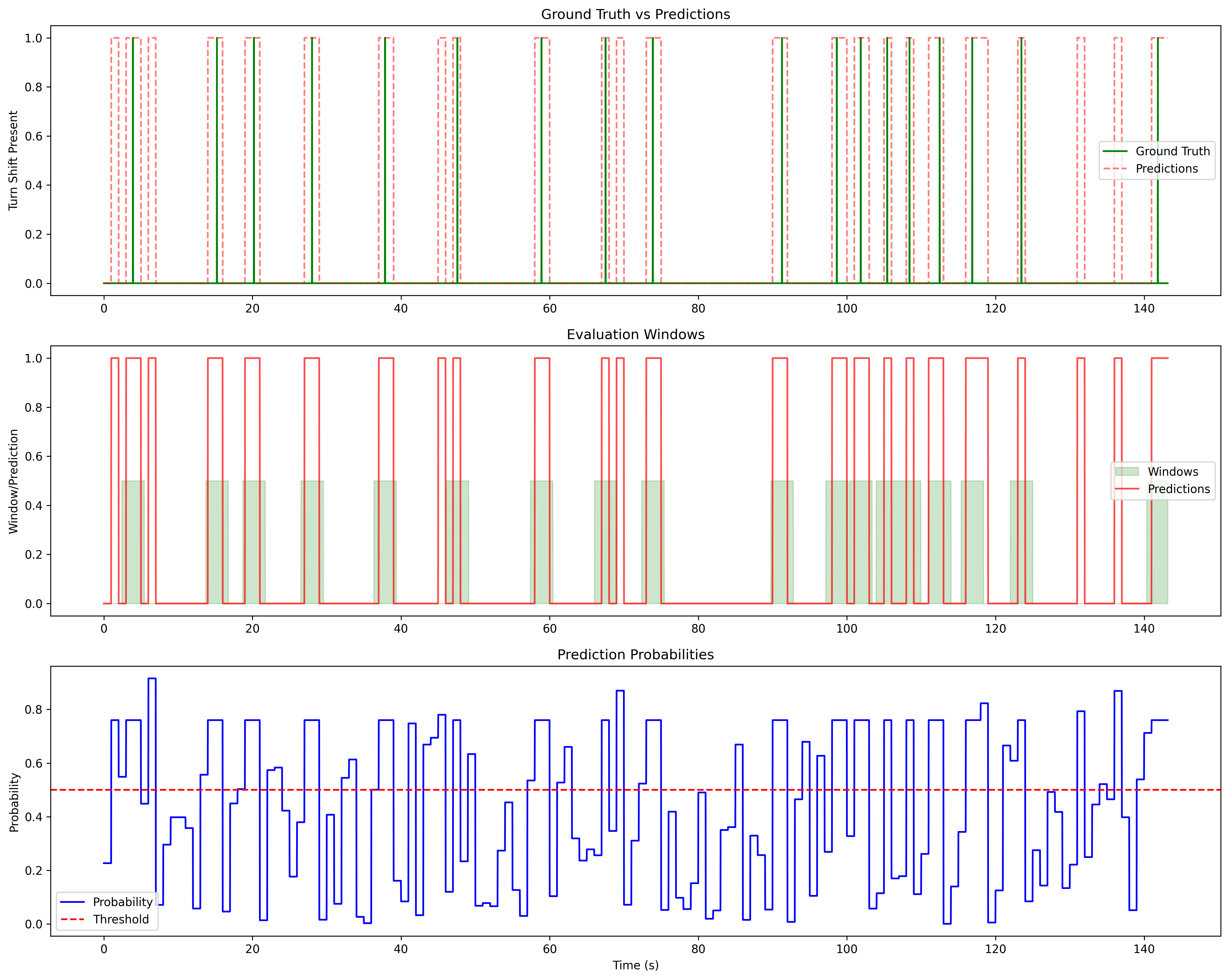}
    \caption{Enhanced prompt inference with improved temporal alignment}
    \label{fig:prompt2_example}
\end{figure}

\begin{figure}[t]
    \centering
    \includegraphics[width=\columnwidth]{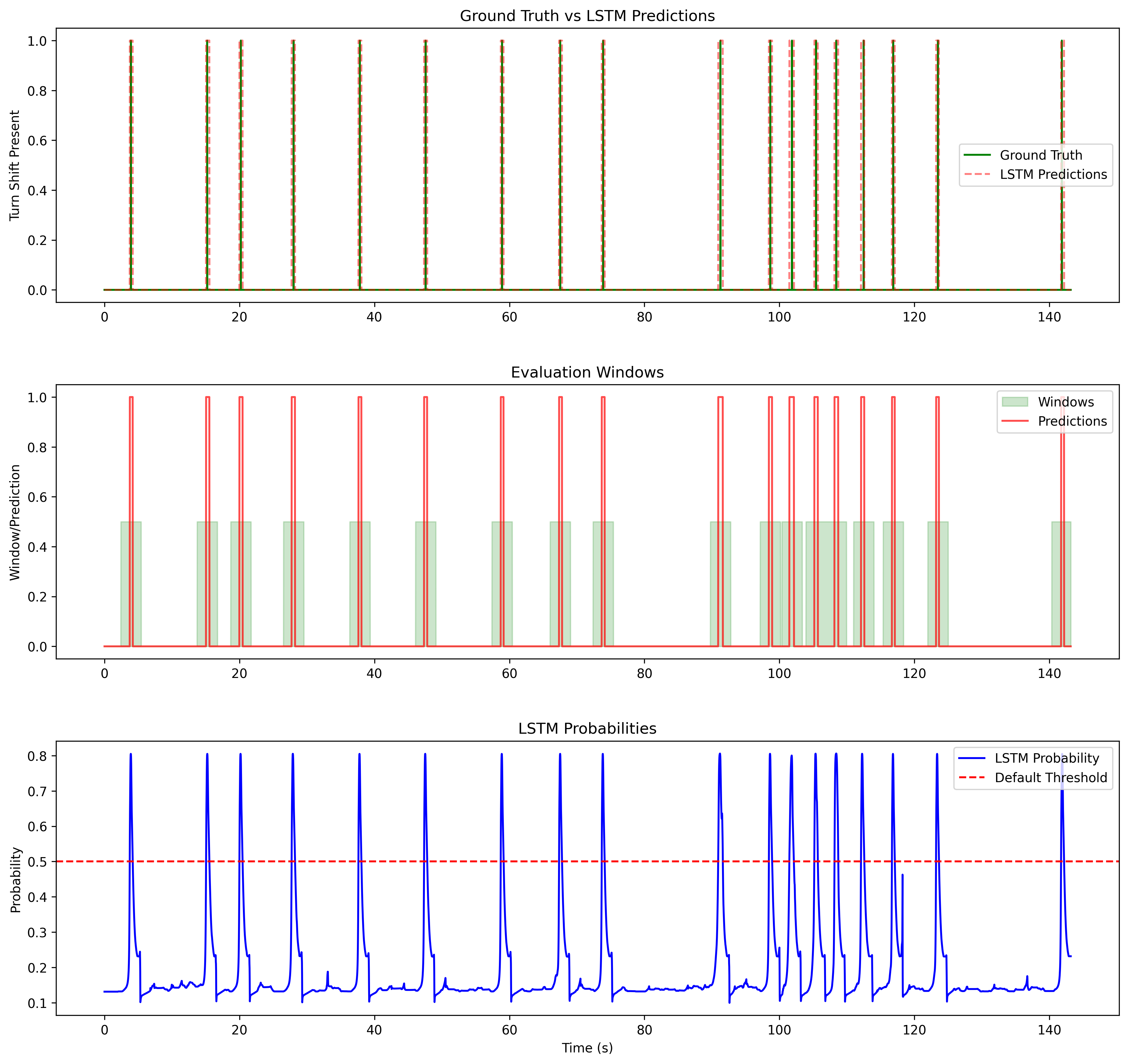}
    \caption{LSTM ensemble inference demonstrating more precise turn-taking predictions}
    \label{fig:lstm_example}
\end{figure}

To compare the real-world performance of our different approaches, we analyze predictions from three model variants on the same conversation segment, visualized in Figures \cref{fig:VAP_example,fig:prompt1_example,fig:prompt2_example,fig:lstm_example}. Each figure presents three key components: ground truth versus model predictions (top), evaluation windows (middle), and prediction probabilities over time (bottom). This visualization structure allows us to directly observe how the basic prompt, enhanced prompt, and LSTM ensemble handle identical turn-taking scenarios. In particular, we can trace the progressive improvements in prediction stability and temporal alignment across these approaches, from the basic prompt's less stable predictions to the LSTM ensemble's more precise turn-taking detection.

\subsection{Within-Turn vs Turn-Final TRP Detection}
Our results reveal a fundamental challenge in TRP detection that varies by context:

\textbf{Within-Turn Detection (ICC Dataset):}\\
Within-turn prediction   means predicting a potential turn change at any point during a speaker's turn, not just at the very end, by looking at earlier linguistic or prosodic features within the ongoing utterance.

\begin{itemize}
    \item Both VAP and Llama struggle significantly with within-turn TRP detection, as evidenced by low precision scores (0.0232 and 0.0259 respectively)
    \item The high recall (0.8862) but extremely low precision (0.0228) for Prompt 1 suggests over-prediction of TRPs, indicating difficulty in distinguishing subtle within-turn transition points
    \item Even the LSTM ensemble, while achieving the highest balanced accuracy (53.80\%) on ICC, shows limited improvement over baselines, suggesting the fundamental difficulty of this task
\end{itemize}

\textbf{Turn-Final Detection (CCPE Dataset):}\\
Turn-final prediction refers to predicting when a speaker is about to finish their turn by analyzing cues near the end of their utterance.

\begin{itemize}
    \item VAP demonstrates strong performance (87.30\% accuracy) on turn-final prediction, suggesting audio features are particularly effective for detecting complete turn boundaries
    \item Llama's improved performance with Prompt 2 (79.26\% accuracy) indicates that linguistic cues are more reliable for turn-final prediction
    \item The LSTM's superior performance (93.20\% accuracy) suggests that turn-final predictions benefit significantly from the combination of audio and linguistic features
\end{itemize}

\subsection{Impact of Prompt Engineering}
Our prompt analysis reveals nuanced effects of different formulation strategies:

\textbf{Prompt Evolution:}
\begin{itemize}
    \item \textbf{Technical vs. Conversational Framing:}
        \begin{itemize}
            \item Prompt 1's technical focus on TRPs yielded poor precision (0.5197) on CCPE
            \item Prompt 2's conversational framing improved precision substantially (0.8300), suggesting LLMs better understand turn-taking when framed as dialogue participation
        \end{itemize}
    \item \textbf{Few-Shot Learning Effects:}
        \begin{itemize}
            \item Prompt 3's perfect recall (1.000) but poor precision indicates that examples biased the model toward over-prediction
            \item This suggests that few-shot learning might introduce unintended biases in turn-taking prediction
        \end{itemize}
\end{itemize}

\subsection{Training Dynamics and Model Behavior}
The LSTM training progression (Figure \ref{fig:lstm_training}) reveals important insights about model learning. Examining the \textbf{Learning Characteristics}, we observe initial rapid improvement followed by stabilization, suggesting efficient feature integration. The convergence pattern of sensitivity/specificity curves indicates balanced learning despite dataset imbalance, while the positive prediction ratio stabilization near 0.5 suggests successful modeling of natural turn-taking frequency.
Analysis of prediction visualizations (Figure\cref{fig:VAP_example,fig:prompt1_example,fig:prompt2_example,fig:lstm_example}) demonstrates evolving model capabilities in \textbf{Temporal Prediction Patterns}. Regarding probability trajectory evolution, the basic prompt shows high-frequency fluctuations in prediction probabilities, while the enhanced prompt demonstrates smoother transitions but some temporal misalignment. The LSTM ensemble achieves sharp, well-defined probability peaks aligned with true turn boundaries, representing a significant improvement over earlier approaches.
In the context of false prediction analysis, false positives often occur near true TRPs, suggesting detection of valid but premature turn-taking opportunities. False negatives are more common in within-turn positions, indicating conservative prediction in ambiguous cases. This pattern demonstrates the model's tendency to err on the side of caution when faced with uncertainty in turn-taking situations.

\subsection{Feature Integration Analysis}
Analysis of different feature integration approaches reveals several key insights:

\textbf{Base Features:}
\begin{itemize}
    \item Audio features from VAP show strong performance on turn-final prediction (87.30\% accuracy on CCPE)
    \item Text features through Llama demonstrate sensitivity to prompt design, achieving best performance with conversational framing (79.26\% accuracy)
    \item ASR quality (Whisper small model, 7.4\% WER) impacts the reliability of text features, potentially limiting Llama's performance
\end{itemize}

\textbf{Integration Strategies:}
\begin{itemize}
    \item Linear regression ensemble (78.20\% accuracy on CCPE) demonstrates basic feature combination
    \item Prompt ensemble improves over base Llama (82.93\% vs 79.26\%) through VAP confidence integration
    \item LSTM ensemble achieves superior performance (93.20\%) through temporal modeling of feature interactions
\end{itemize}

\subsection{Architectural Insights}
Analysis of different architectural approaches reveals their strengths and limitations. For \textbf{Linear Regression}, the simple linear combination proves insufficient for complex feature relationships. This approach demonstrates limited ability to capture temporal dependencies, though it provides a useful baseline for evaluating feature integration effectiveness.
The \textbf{Prompt-based Architecture} successfully incorporates audio confidence into language model reasoning and maintains model interpretability through explicit prompt engineering. However, this approach shows notable limitations in handling temporal aspects of turn-taking dynamics.
In contrast, the \textbf{LSTM-based Architecture} effectively captures temporal dependencies in both audio and text features. Its multi-head attention mechanism enables complex feature interactions, while demonstrating robustness to ASR errors through integrated feature processing.
Regarding \textbf{Efficiency Considerations}, all architectures maintain real-time processing capability, presenting a clear trade-off between architectural complexity and performance improvement. The LSTM's additional computational cost, with a Real-Time Factor (RTF) of 0.0249, is justified by its significant performance gains over simpler architectures.

\section{Conclusion}
This study presents a comprehensive analysis of turn-taking prediction approaches, comparing audio-based, text-based, and ensemble methods across different conversational contexts. Our analysis reveals several key findings. The context dependency of prediction accuracy is evident, with turn-final prediction achieving high accuracy (up to 93.20\% with LSTM ensemble on CCPE), while within-turn prediction remains challenging across all approaches. In terms of modality integration, the successful combination of audio and text features through LSTM-based temporal modeling significantly outperforms single-modality approaches. We found that language model performance heavily depends on task framing, with conversational prompts outperforming technical descriptions. Importantly, all tested approaches maintain real-time processing capability (RTF < 0.05), making them suitable for practical applications.
\subsection{Practical Implications}
Our findings have direct implications for real-world applications. For structured dialogue with clear turn boundaries, VAP alone may be sufficient, while more complex scenarios benefit from ensemble approaches, though within-turn prediction remains challenging. From a deployment perspective, all models support real-time deployment with varying computational requirements. Organizations can optimize model choice based on specific accuracy needs and resource constraints, though careful threshold tuning remains crucial for optimal performance.
\subsection{Limitations and Future Work}
Several challenges and opportunities remain for future research. Within-turn TRP detection requires more sophisticated modeling approaches, and ASR errors (7.4\% WER) may limit text-based prediction accuracy. The relationship between linguistic and acoustic features in turn-taking warrants deeper investigation. Future improvements could be achieved through better prompt engineering and language model integration, specialized approaches for different turn-taking scenarios, and more sophisticated ensemble architectures.

\paragraph{Source Code:}
\href{https://github.com/HarryJeon24/LlaVap}{LlaVAP}

\appendix
\section{Author Contributions}
Harry Jeon and Rayvant Sahni contributed to this work. H.J. designed and implemented the experimental framework, developed the model architectures, conducted all computational experiments, and performed data analysis. R.S. led the manuscript preparation and writing, synthesized the experimental results into coherent findings, and structured the technical documentation. Both authors reviewed and approved the final manuscript.

%\bibliographystyle{acl_natbib}
\bibliography{custom}
\cleardoublepage

\end{document}